\documentclass[runningheads]{llncs}
\usepackage{graphicx}

\usepackage{amsmath}    
\usepackage{amssymb}    

\usepackage{multirow} 
\usepackage{booktabs}

\usepackage[table,xcdraw,usenames,dvipsnames,table]{xcolor}

\begin{document}
\title{Soft Tissue Sarcoma Co-Segmentation in Combined MRI and PET/CT Data}
\titlerunning{Sarcoma Co-Segmentation in Combined MRI and PET/CT Data}

\author{Theresa Neubauer\inst{1} \and
	Maria Wimmer\inst{1} \and 
	Astrid Berg\inst{1} \and 
	David Major\inst{1} \and 
	Dimitrios Lenis\inst{1} \and 
	Thomas Beyer\inst{2} \and
	Jelena Saponjski\inst{3} \and 
	Katja B\"uhler\inst{1}
}


\authorrunning{T. Neubauer et al.}

\institute{VRVis Zentrum f\"ur Virtual Reality und Visualisierung Forschungs-GmbH, \\Vienna, Austria \\ \email{mwimmer@vrvis.at} \and 
QIMP Team, Center for Medical Physics and Biomedical Engineering, Medical University of Vienna, Austria \and
Center for Nuclear Medicine, Clinical Center of Serbia, Belgrade, Serbia
\\}

\maketitle              

\begin{abstract}
Tumor segmentation in multimodal medical images has seen a growing trend towards deep learning based methods.
Typically, studies dealing with this topic fuse multimodal image data to improve the tumor segmentation contour for a single imaging modality. 
However, they do not take into account that tumor characteristics are emphasized differently by each modality, which affects the tumor delineation. Thus, the tumor segmentation is modality- and task-dependent. This is especially the case for soft tissue sarcomas, where, due to necrotic tumor tissue, the segmentation differs vastly. 
Closing this gap, we develop a modality-specific sarcoma segmentation model that utilizes multimodal image data to improve the tumor delineation on each individual modality.
We propose a simultaneous co-segmentation method, which enables multimodal feature learning through modality-specific encoder and decoder branches, and the use of resource-efficient densely connected convolutional layers. 
We further conduct experiments to analyze how different input modalities and encoder-decoder fusion strategies affect the segmentation result.
We demonstrate the effectiveness of our approach on public soft tissue sarcoma data, which comprises MRI (T1 and T2 sequence) and PET/CT scans. The results show that our multimodal co-segmentation model provides better modality-specific tumor segmentation than models using only the PET or MRI (T1 and T2) scan as input. 

\end{abstract}

\section{Introduction}
In cancer therapy, automatic tumor segmentation supports healthcare professionals as it provides a fast quantitative description of the tumor volume and location. To analyze soft tissue sarcomas in more detail, usually, complementing imaging modalities are used to depict the tumor from an anatomical or physiological perspective, such as Magnetic Resonance Imaging (MRI), Computed Tomography (CT), or Positron Emission Tomography (PET). These modalities show different characteristics of the tumor tissue and thus provide valuable complementary information.  
However, depending on the imaging modality and clinical indication, the segmentation contour may look different for the same tumor. 

\begin{figure}
	[!htbp]
	\center
	\includegraphics[width=1\textwidth]{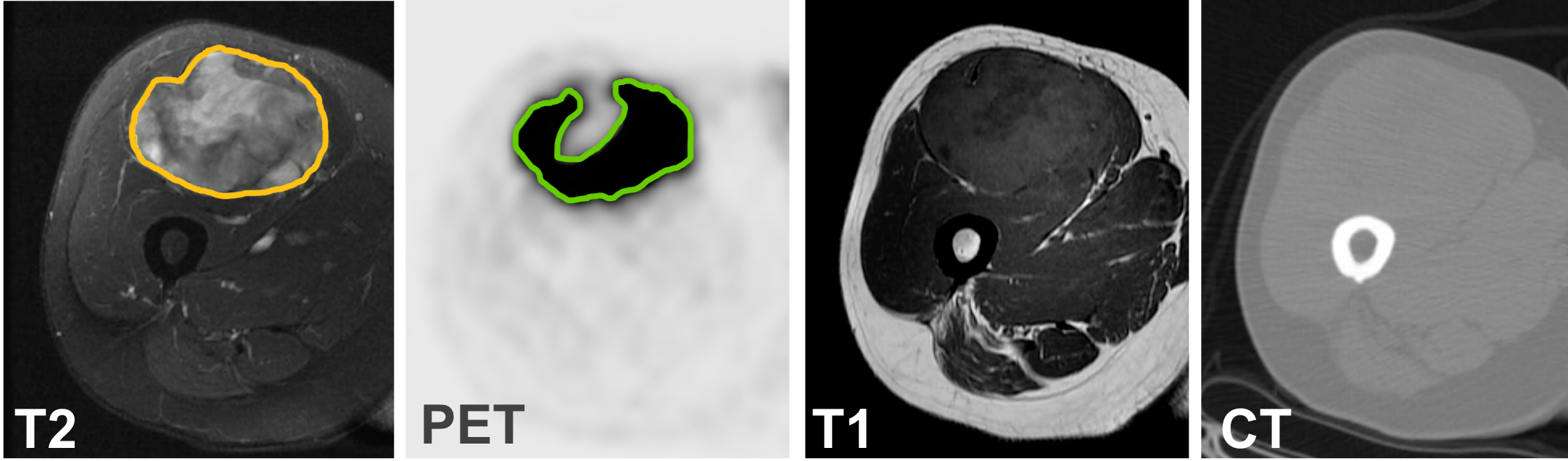}
	\caption{Depending on the modality and the clinical intent, the segmentation for soft tissue sarcomas on the MRI T2 scan (yellow contour) and the PET scan (green contour) may look different. Figure best viewed in color.} 
	\label{fig:multimodal_segmentation_challenge}
\end{figure}

Soft tissue sarcomas are malignant tumors that originate from various tissues, including muscular tissue, connective tissue, and nervous tissue. They predominantly occur in the extremities. Due to their large size, soft tissue sarcomas tend to form necrotic tumor areas. In MRI scans, necrosis is considered part of the tumor, but it is not visible on the PET scan as the necrosis is no longer metabolically active. 
Fig.~\ref{fig:multimodal_segmentation_challenge} demonstrates the challenge of multimodal segmentation for soft tissue sarcomas on PET and MRI scans.

Deep learning based multimodal tumor segmentation methods have been proposed, e.g. for brain tumor segmentation on multi-sequence MRIs \cite{ref_isensee,ref_myronenko} or lung tumor segmentation on PET/CTs \cite{ref_kumar,ref_zhong}. Current state-of-the-art networks are inspired by fully convolutional neural networks (FCNs), whereby different ways to incorporate the complementary information of multimodal image data have been presented. 
These multimodal segmentation studies report a better segmentation result compared to  models using monomodal images. However, the main limitation of these studies is that one modality is set as the segmentation target for the final contour and thus only one modality-specific tumor volume is obtained. Contrary, in cancer therapy there are different clinical routines, which require a set of modality-specific tumor delineations from the input data. 

To solve this problem for sarcomas, we aim to \textit{simultaneously co-segment} selected modality-specific tumor volumes from the given input modalities.
To the best of our knowledge, there is only the study of Zhong et al.~\cite{ref_zhong}, which investigates tumor co-segmentation with deep learning. They perform lung tumor segmentation on PET/CT scans, co-segmenting the modality-specific tumor in both the CT and PET scan. However, their use of two connected 3D U-Nets (one per modality), results in a very large model with more than 30M parameters.

Therefore, we introduce a resource-efficient, multimodal network for sarcoma co-segmentation, which allows the network to simultaneously segment several modality-specific tumor volumes on a subset of the input modalities. 
Our model benefits from (1) modality-specific encoders and decoders for multimodal feature learning, and (2) dense blocks for efficient feature re-use. We demonstrate the effectiveness of our method on public soft-tissue sarcoma data~\cite{ref_tcia,ref_dataset,ref_dataset_paper} and extensively evaluate the influence of MRI and PET/CT data for co-segmentation.

\section{Method}
\label{section:network_architecture}

For each patient $i$, $i=1,\ldots,n$, let $\mathcal{I}_i$ be a set of medical images of fixed modalities corresponding to this patient, i.e. $\mathcal{I}_i := \{I_i^m\}_{i,m}$ with $I_i^m$ an image of patient $i$ and modality $m \in \{T1,T2,CT,PET\}$. For every $\mathcal{I}_i$, we define the set of corresponding ground truth segmentation masks $\mathcal{M}_i := \{M_i^{m'}\}_{i,m'}$ where $m' \in \{T2,PET\}$. 
We then seek for a co-segmentation network that is capable of estimating the given ground truth masks $\mathcal{M}_i$, given a chosen subset of input modalities.
Our proposed model is inspired by the popular U-Net~\cite{ref_ronneberger} architecture, and the work of J\'{e}gou et al.~\cite{ref_jegou}, who extended the DenseNet~\cite{ref_huang} for the task of semantic segmentation. Fig.~\ref{fig:network} gives an overview of our model, which comprises the following main parts:

\begin{figure}
	[!htbp]
	\center
	\includegraphics[width=1\textwidth]{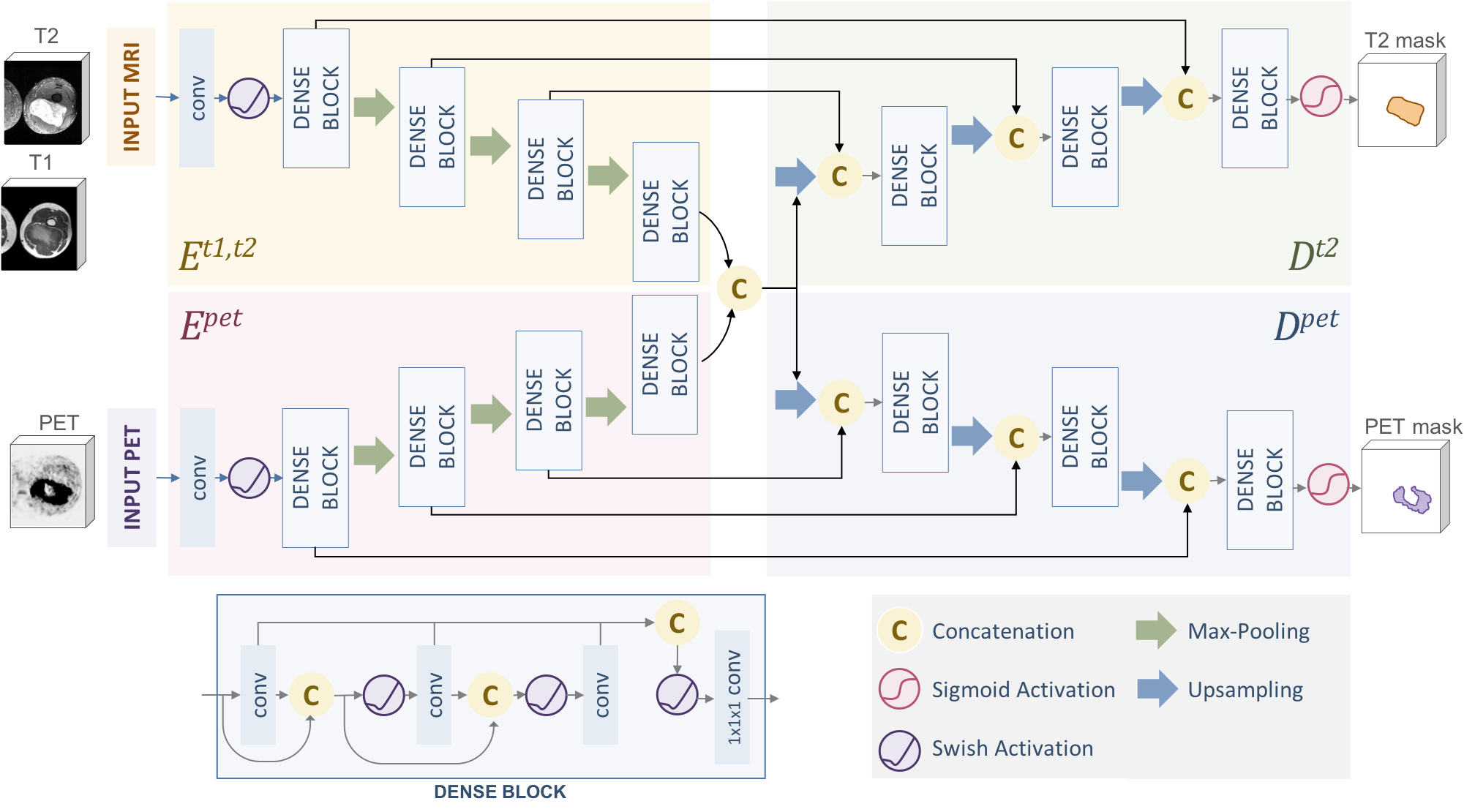}
	\caption{We use two separated encoder branches $E^{t1,t2}$ and $E^{pet}$ for modality-specific feature extraction and pass the concatenated latent representation to both decoders $D^{t2}$ and $D^{pet}$ for efficient segmentation of both tumor contours. Best viewed in color.} 
	\label{fig:network}
\end{figure}

\paragraph{\textbf{Modality-Specific Encoder and Decoder }}

We use two different encoder branches $E^{t1,t2}$ and $E^{pet}$ for MRI and PET data, respectively, to extract features for each target modality separately. 
In the MRI branch, we additionally use the T1 scan as a supporting modality to improve the feature extraction of the target T2 scan. 
The separation of the modality types in the encoder part is inspired by prior work on multimodal segmentation models \cite{ref_zhou}. Firstly, studies with multi-sequence MRIs have shown that input-level fusion leads to a significant improvement in model performance \cite{ref_isensee,ref_myronenko}. Secondly, for studies dealing with complementary modalities such as PET/CT, modality-specific encoder branches are widely used \cite{ref_kumar,ref_zhong}.

Each encoder, $E^{t1,t2}$ and $E^{pet}$, consists of an initial $3 \times 3 \times 3$ convolution layer with 48 filters, followed by four dense blocks. After each block, the resulting feature map is then downsampled using a max-pooling layer with pooling size $ 2 \times 2  \times 2$ and stride 2, which halves the resolution of the feature maps. To account for the low spatial resolution of the z-axis of the MRI scans, we propose to perform $2 \times 2 \times 1$ pooling after the second dense block instead. 

We concatenate the latent representation of $E^{t1,t2}$ and $E^{pet}$ and pass the feature maps to both decoders $D^{t2}$ and $D^{pet}$. Each dense block in each decoder $D^{t2}$ and $D^{pet}$ receives the feature maps of the dense block at the same resolution level from the corresponding encoder $E^{t1,t2}$ and $E^{pet}$, respectively. In the following we refer to our proposed model as $E^{t1,t2}E^{pet}\textit{-}D^{t2}D^{pet}$.

\paragraph{\textbf{Dense Blocks }}

Each dense block consists of three repeated $ 3 \times 3  \times 3$ convolution layers and Swish~\cite{ref_swish} activations. This iterative concatenation of feature maps leads to feature re-use, which in turn reduces the number of parameters \cite{ref_huang}. The number of filters of all convolution layers in a block is increased with each block level, learning 12, 28, 44, or 60 filters, respectively. In contrast to J\'{e}gou et al.~\cite{ref_jegou}, we removed the batch normalization layers, since we use a batch size of one. We also removed the dropout layers, because they did not lead to performance improvements. At the end of the dense block, the feature maps of all convolution layers are then concatenated and reduced by a factor of 0.5 using a $ 1 \times 1  \times 1$ convolution layer to further decrease the number of model parameters. 

\paragraph{\textbf{Loss function }}
To account for both tumor masks in our co-segmentation model during training, we calculate the dice losses individually for each mask in $\mathcal{M}_i$ and combine them as follows:

\begin{equation}
\mathcal{L} = -\sum_{m' \in \{T2,PET\}}^{} \dfrac{2 \mid M^{m'} \cap P^{m'} \mid + \:\epsilon}{\mid M^{m'} \mid + \mid P^{m'} \mid + \:\epsilon} 
\label{eq:dice_loss}
\end{equation}

whereby $ M^{m'}$ and $ P^{m'}$ denote the voxel set of the ground truth volume $M^{m'}$ and the predicted volume $P^{m'}$ belonging to modality $m'~\in~\{T2,PET\}$. The parameter $\epsilon$ is added to avoid numerical instabilities.

\paragraph{\textbf{Variant: Shared Decoder }}
We further introduce a lightweight variant of our model which uses only one shared decoder $D^{t2,pet}$. Here, each dense block receives the multiplied feature maps from the $E^{t1,t2}$ and $E^{pet}$ encoder block at the same level. The fusion of feature maps by multiplication is intended to emphasize the overlapping position of the two masks. However, the feature maps of the first encoder blocks are fused by concatenation to allow for modality-specific differences in the segmentation masks. The last layer of the decoder has two output channels: one for the MRI mask $M_i^{t2}$ and one for the PET mask $M_i^{pet}$. We compare both models in Section~\ref{section:results}.

\section{Experimental Setup}
\subsection{Dataset and Pre-processing}
We evaluate our method on the soft tissue sarcoma dataset~\cite{ref_dataset,ref_dataset_paper}, which is publicly available at The Cancer Imaging Archive~\cite{ref_tcia}. 
The highly heterogeneous dataset comprises 51 patients with sarcomas in the extremities, with the data coming from different sites and scanners. For each patient, four different imaging modalities have been acquired: two paired MRI (T1 and T2) scans and a PET/CT scan. The MRI and PET/CT exams were acquired on different days, resulting in changed body positions as well as anatomical variations. The dataset already includes tumor annotations, which are delineated on the T2 scans. In addition, an experienced nuclear physician delineated the tumor contours for our study on the PET scan for radiotherapy treatment. 
We pre-processed the dataset as follows:
\begin{itemize}
	\item \textbf{Co-registration}: We followed Leibfarth et al. \cite{ref_leibfarth} for multimodal intra-patient registration and registered the PET/CT scan with the corresponding PET contour on the T2 scan.  
	\item \textbf{Resampling}: The in-plane pixel resolution was resampled to $ 0.75 \times 0.75$~mm using B-Spline interpolation, while the slice distance was kept at the original distance of the T2 scan to avoid resampling artifacts due to the low spatial resolution.
	\item \textbf{Crop images}: We focus on patients with tumors in their legs and cropped all scans to the leg region, resulting in 39 patients. The cropped scans have varying sizes ($210 \times 210$ to $ 600 \times 660$) and slice numbers (15 to 49).
	\item \textbf{Modality-dependent intensity normalization}: We applied z-score normalization to the T1 and T2 scans. The PET scans were normalized by a transformation to standard uptake values using body-weight correction.
\end{itemize}

\subsection{Network Training}
We randomly divide the 39 patients into five distinct sets and perform 5-fold cross-validation. We increase the efficiency of the training using 3D patches of size $ 256 \times 256 \times 16 $, which are randomly extracted from the image while ensuring that tumor tissue is visible on every patch. To avoid overfitting and account for the small number of training samples, we perform the following data augmentation strategies: scaling, rotation, mirroring, and elastic transformations.
We train our network using the loss function Eq.~\ref{eq:dice_loss} and the Adam optimizer with a batch size of one. We start with an initial learning rate of $ 1e^{-4} $, which is reduced by a factor of 0.5 if the validation loss has not decreased for eight epochs. All convolutional kernels are initialized with he\_normal.
The models are implemented using Keras with Tensorflow backend and trained on an NVIDIA Titan RTX GPU (24 GB RAM).

\subsection{Evaluation measures}
The segmentation performance is measured calculating the overlap-based dice similarity coefficient (DSC) and distance-based average symmetric surface distance (ASSD) for each predicted mask $ P $ of modality $m'$ and its corresponding ground truth mask $M \in \mathcal{M}_i $. Formally: 
\begin{equation}
DSC^{m'} (M, P) = \frac{2 \mid M \cap P \mid}{\mid M \mid + \mid P\mid}
\end{equation}
\begin{equation}
ASSD^{m'}(M,P) = \dfrac{\sum_{g_{k} \in M}d(g_{k},M) + \sum_{p_{k} \in P} d(p_{k},P)}{\mid M \mid + \mid P \mid }
\end{equation}
whereby $g_{k} \in M$ and $p_{k} \in P$ denote a voxel in the ground truth volume $ M $ and  predicted volume $ P $, respectively. The Euclidean distance $ d(g_{k},P) $ is calculated between voxel $g_{k}$ and the closest voxel in $P$.

\section{Results and Discussion}
\label{section:results}

\begin{figure}
	[!htbp]
	\includegraphics[width=1\textwidth]{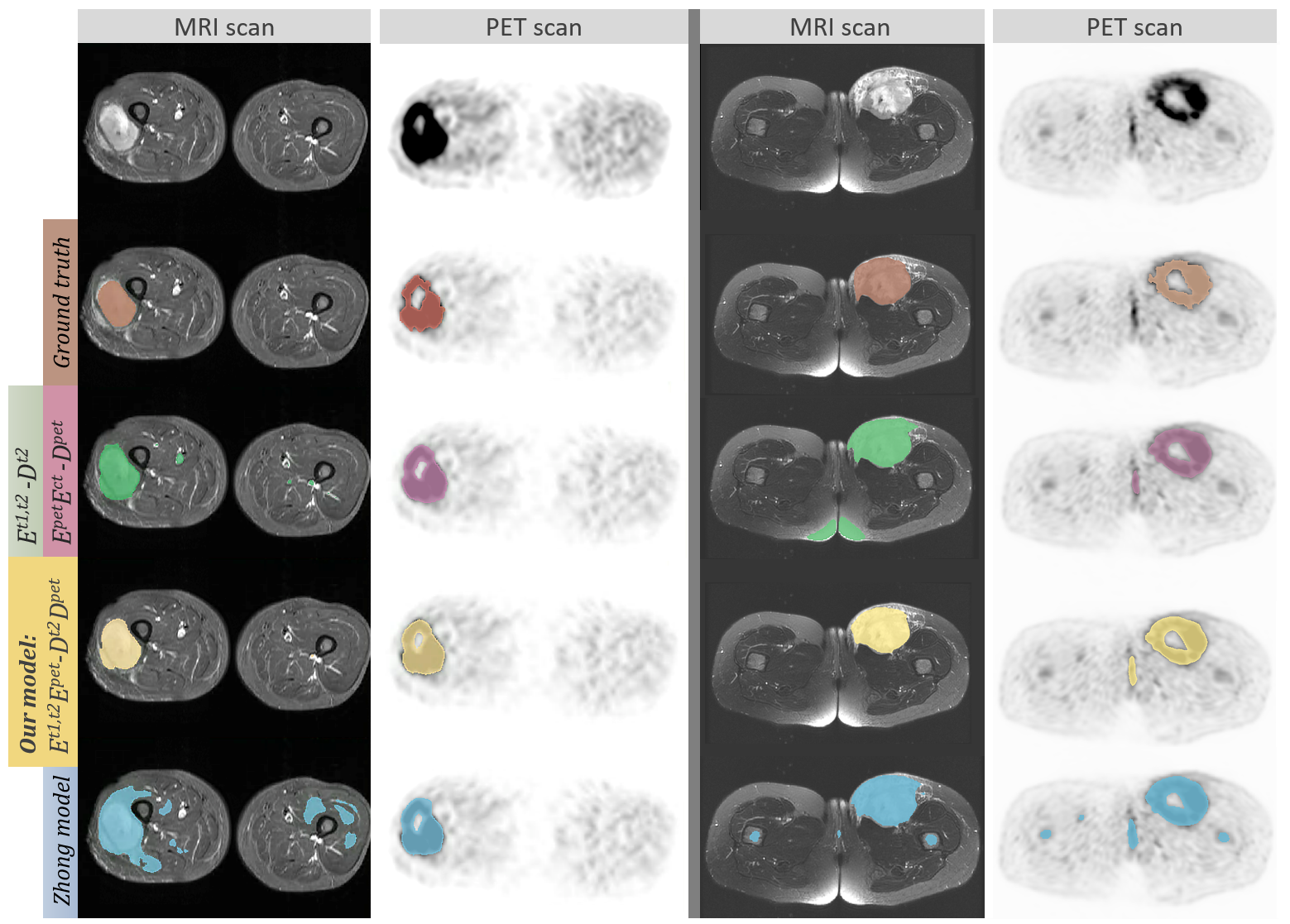}
	\caption{Visual segmentation results of compared models on T2 and PET scan pairs. The presented samples confirm the trend from Table \ref{tab:result_scores}: The variation of the predicted T2 masks is higher between different models, while the impact for the PET segmentations is less apparent. Best viewed in color.
	} 
	\label{fig:result_images}
\end{figure}

We compare the performance of our proposed network $E^{t1,t2}E^{pet}\textit{-}D^{t2}D^{pet}$ with different baseline models. These experiments demonstrate the influence of varying sets of input modalities as well as modality-specific encoder/decoder designs for our model. Table \ref{tab:result_scores} summarizes mean DSC (in \%) and ASSD (in mm) for T2 and PET segmentation separately. Visual results are shown in Fig.~\ref{fig:result_images}. To compare our approach to the state-of-the-art, we implement the model by Zhong et al.~\cite{ref_zhong} using two parallel U-Nets: one for the T2 scan and one for the PET scan yielding the segmentation masks for the T2 and PET scan simultaneously. 
We followed the proposed implementation details. 
However, to allow for a fair comparison, we changed the patch size to our settings. Additionally, we adapt our z-axis pooling approach to the model of Zhong et al. and name it \textit{Zhong modified}.

\begin{table}[]
	\caption{Performance metrics per model: Mean DSC and ASSD and their standard deviation calculated for the T2 and PET segmentation masks. All results were obtained by running a 5-fold cross-validation. Modalities used in the first encoder branch are denoted by $\bullet$, and the ones in the second encoder branch are denoted by $\circ$.}
	\label{tab:result_scores}
	\begin{tabular}{|l|cccc|cc|cc|}
		\hline
		\multicolumn{1}{|c|}{}                                 & \multicolumn{4}{c|}{\textbf{Input}}                                                                                             & \multicolumn{2}{c|}{\textbf{Mean DSC (\%)}}                                           & \multicolumn{2}{c|}{\textbf{ASSD (mm)}}                                                     \\
		\multicolumn{1}{|c|}{\multirow{-2}{*}{\textbf{Model}}} & T1                                & T2                                & PET     & CT                                            & T2                                                           & PET                    & T2                                                          & PET                           \\ \hline
		$E^{t1,t2}E^{pet}\textit{-}D^{t2}D^{pet}$                                      & $\bullet$ & $\bullet$ & $\circ$ & \multicolumn{1}{l|}{} & \multicolumn{1}{c|}{\textbf{$\:$\textbf{77.2}$\pm$\textbf{16.5}$\:\:$}} & $\:74.6\pm19.0\:\:$    & \multicolumn{1}{c|}{\textbf{$\:$ \textbf{3.8}$\pm$\textbf{5.3}$\:\:$}} & $\:4.5\pm6.2\:$               \\
		$E^{t1,t2}E^{pet}\textit{-}D^{t2,pet}$                                         &  $\bullet$ & $\bullet$ & $\circ$                         & \multicolumn{1}{l|}{} & \multicolumn{1}{c|}{$75.3\pm17.2$}                           & $74.2\pm19.9$          & \multicolumn{1}{c|}{$4.5\pm5.3$}                            & $4.3\pm5.4$                   \\
		$E^{t1,t2}E^{pet}\textit{-}D^{t2}$                                             & $\bullet$ & $\bullet$ & $\circ$                         & \multicolumn{1}{l|}{} & \multicolumn{1}{c|}{$76.5\pm16.6$}                           & .                      & \multicolumn{1}{c|}{$3.9\pm4.9$}                            & .                             \\
		$E^{t1,t2}E^{pet}\textit{-}D^{pet}$                                            & $\bullet$ & $\bullet$ & $\circ$                         & \multicolumn{1}{l|}{} & \multicolumn{1}{c|}{.}                                       & $74.9\pm16.1$          & \multicolumn{1}{c|}{.}                                      & $4.3\pm5.1$                   \\
		$E^{t2}\textit{-}D^{t2}$                                                       &                                   & $\bullet$                         &                                 &                       & \multicolumn{1}{c|}{$65.6\pm24.0$}                           & .                      & \multicolumn{1}{c|}{$10.2\pm10.9$}                          & .                             \\
		$E^{t1,t2}\textit{-}D^{t2}$                                                    & $\bullet$ & $\bullet$ &                                 &                       & \multicolumn{1}{c|}{$71.0\pm23.8$}                           & .                      & \multicolumn{1}{c|}{$6.5\pm8.2$}                            & .                             \\
		$E^{t1}E^{t2}\textit{-}D^{t2}$                                                 & $\bullet$ & $\circ$   &        &                                               & \multicolumn{1}{c|}{$68.3\pm20.0$}                           & .                      & \multicolumn{1}{c|}{$7.9\pm8.9$}                            & .                             \\
		$E^{pet}\textit{-}D^{pet}$                                                     &                                   &           & $\bullet$                       &                                               & \multicolumn{1}{c|}{.}                                       & $74.3\pm18.8$          & \multicolumn{1}{c|}{.}                                      & $4.9\pm6.4$                   \\
		$E^{pet,ct}\textit{-}D^{pet}$                                                  &                                   &                                   & $\bullet$                       & $\bullet$             & \multicolumn{1}{c|}{.}                                       & $74.4\pm21.6$          & \multicolumn{1}{c|}{.}                                      & $\:\:5.5\pm14.1$              \\
		$E^{pet}E^{ct}\textit{-}D^{pet}$                                               &                                   &                                   & $\bullet$                       & $\circ$               & \multicolumn{1}{c|}{.}                                       & \textbf{76.1}$\pm$\textbf{16.0} & \multicolumn{1}{c|}{.}                                      & $3.7\pm4.1$                   \\
		$E^{t2}E^{pet}\textit{-}D^{t2}D^{pet}$                                         &                                   & $\bullet$                         & $\circ$ &                                               & \multicolumn{1}{c|}{$72.1\pm19.8$}                           & $73.5\pm20.0$          & \multicolumn{1}{c|}{$4.9\pm5.3$}                            & $4.8\pm6.2$                   \\
		$E^{t2}E^{pet}\textit{-}D^{t2,pet}$                                            &                                   & $\bullet$                         & $\circ$ &                                               & \multicolumn{1}{c|}{$71.6\pm19.4$}                           & $73.2\pm19.5$          & \multicolumn{1}{c|}{$5.6\pm6.0$}                            & $4.2\pm5.1$                   \\
		Zhong~\cite{ref_zhong}                                                         &                                   & $\bullet$                         & $\circ$ &                                               & \multicolumn{1}{c|}{$72.4\pm20.0$}                           & $74.1\pm19.4$ & \multicolumn{1}{c|}{$6.7\pm8.8$}                            & \textbf{$4.0\pm4.9$}          \\
		Zhong modified                                                                 &                                   & $\bullet$ & $\circ$                         &                                               & \multicolumn{1}{c|}{$75.6\pm16.2$}                           & $75.2\pm17.2$ & \multicolumn{1}{c|}{$4.8\pm6.5$}                            & \textbf{3.6}$\pm$\textbf{4.3} \\ \hline
	\end{tabular}
\end{table}

\paragraph{\textbf{Single modality mask prediction:}}

When comparing the scores for the prediction of $M_i^{t2}$ only, we found that the lowest results are achieved when only using T2 as input. The performance increases when incorporating both T1 and T2 in the encoders, whereby the best results are obtained with a shared encoder used in model $E^{t1,t2}\textit{-}D^{t2}$. These results confirm our choice for the shared MRI encoder $E^{t1,t2}$ of our proposed model.
In contrast, a single PET modality is sufficient to achieve a good PET segmentation $M_i^{pet}$, as shown for model $E^{pet}\textit{-}D^{pet}$.
We further observed, that adding a separate encoder $E^{ct}$ to the model resulted in the highest performance increase, yielding the best scores for predicting $M_i^{pet}$ overall (76.1\% $\pm$ 16.0\% DSC, 3.7~mm~$\pm$~4.1~mm ASSD).

\paragraph{\textbf{Encoder/Decoder design:}}

The results in Table~\ref{tab:result_scores} suggest that the segmentation performance benefits from modality-specific encoders that separate anatomical and functional modalities. 
Comparing the models with shared $D^{t2,pet}$ and separate decoders $D^{t2}D^{pet}$, we report lower DSC scores when using the proposed shared decoder variant of our model. The performance impact is higher for T2, which is also reflected by DSC and ASSD scores.

\paragraph{\textbf{Co-Segmentation:}}

Looking at the sarcoma co-segmentation models, we observe that the tumor delineation on MRI T2 scans benefits from the feature co-learning with PET. This is reflected by the best overall scores (77.2\% $\pm$ 16.5\% DSC, 3.8~mm~$\pm$~5.3~mm ASSD) obtained with our proposed model $E^{t1,t2}E^{pet}\textit{-}D^{t2}D^{pet}$. The same is observed with model $E^{t1,t2}E^{pet}\textit{-}D^{t2}$ for predicting only the T2 mask, which gives comparable results to our model.
Model $E^{t1,t2}E^{pet}\textit{-}D^{t2}D^{pet}$ outperforms the method by Zhong et al.~\cite{ref_zhong} and achieves similar DSC and ASSD values to the model \textit{Zhong modified}. However, our $D^{t2,pet}$ models (max. 3.9M) and $D^{t2}D^{pet}$ models (max. 5.5M) require - depending on the encoder - only 10-16\% of the parameters of Zhong et al. (33.7M) and are therefore much more resource-efficient.
When comparing the model of Zhong et al. with \textit{Zhong modified}, it 
is revealing that the adaption of the pooling strategy to the anisotropic data resolution yields notable performance gains.

\section{Conclusion}
In this paper, we proposed a simultaneous co-segmentation model for soft tissue sarcomas, which utilizes densely connected convolutional layers for efficient multimodal feature learning. We performed an extensive evaluation, comparing various ways to incorporate multimodal data (MRI T1 and T2, CT and PET) into our model. We showed that our proposed network outperforms the state-of-the-art method for tumor co-segmentation, yielding better or comparable results for MRI T2 and PET, respectively. Moreover, our proposed co-segmentation architecture and single-modal variants reduce the number of parameters by up to 90\% compared to the concurring method. 
These experiments show (1) improved accuracy when using multimodal data and (2) demonstrate that the choice of input modalities and encoder-decoder architecture is crucial for the segmentation result.

\subsection*{Acknowledgements}
VRVis is funded by BMK, BMDW, Styria, SFG and Vienna Business Agency in the scope of COMET - Competence Centers for Excellent Technologies (854174) which is managed by FFG.

%
%
%
\bibliographystyle{splncs04}
\bibliography{paper6}

\end{document}